\pdfoutput=1
\documentclass[aps,prb,twocolumn,superscriptaddress]{revtex4}

\usepackage{amssymb,amsfonts,amsmath,amsthm}
\usepackage{color,enumerate,graphicx}
\usepackage{hyperref}
\usepackage[normalem]{ulem}

\newcommand{\eps}{\varepsilon}
\newcommand{\R}{\mathbb R}

\newcommand{\m}{\mathbf{m}}
\newcommand{\mpa}{m_\parallel}

\newcommand{\mpe}{\mathbf{m}_\perp}

\newtheorem{theorem}{Theorem}

\begin{document}

\title{Unraveling the role of dipolar versus Dzyaloshinskii-Moriya interaction in stabilizing compact magnetic skyrmions}



\author{ Anne Bernand-Mantel}

\affiliation{Universit\'e de Toulouse, Laboratoire de Physique et
  Chimie des Nano-Objets, UMR 5215 INSA, CNRS, UPS, 135 Avenue de
  Rangueil, F-31077 Toulouse Cedex 4, France}

\email{bernandm@insa-toulouse.fr}

\author{Cyrill B. Muratov}

\affiliation{Department of Mathematical Sciences, New Jersey Institute
  of Technology, Newark, New Jersey 07102, USA}


\author{Thilo M. Simon} 

\affiliation{Department of Mathematical Sciences, New Jersey Institute
  of Technology, Newark, New Jersey 07102, USA}

\affiliation{Institute for Applied Mathematics, University of Bonn,
  Endenicher Allee 60, 53115 Bonn, Germany}

\date{\today}

\begin{abstract}

  We present a theoretical study of compact magnetic skyrmions in
  ferromagnetic films with perpendicular magnetic anisotropy that
  accounts for the full stray field energy in the thin film and low
  interfacial DMI regime. In this regime, the skyrmion profile is
  close to a Belavin-Polyakov profile, which yields analytical
  expressions for the equilibrium skyrmion radius and energy. The
  obtained formulas provide a clear identification of
  Dzyaloshinskii-Moryia and long-range dipolar interactions as two
  physical mechanisms determining skyrmion size and stability.
\end{abstract}

\maketitle

\section{\label{sec:intro}INTRODUCTION}

Magnetic skyrmions are a prime example of topologically nontrivial
spin textures observed in a variety of magnetic materials.  Their
nucleation and annihilation in an otherwise uniformly magnetized
ferromagnet is enabled by the discrete nature of matter.
\cite{romming13,verga14,cortes-ortuno17,hagemeister15} Magnetic
skyrmions emerge when the exchange and anisotropy energies promoting
parallel alignment of spins in a ferromagnet enter in competition with
energies favoring non-collinear alignment of spins such as the
Dzyaloshinskii-Moriya interaction (DMI),\cite{bogdanov89,bogdanov89a} 
the long-range dipolar interaction \cite{yu16,montoya17} or
higher-order exchange interactions.\cite{ivanov90,abanov98,okubo12} In
particular, DMI is at the heart of a large number of magnetic skyrmion
observations in recent years. This antisymmetric exchange interaction
is related to the lack of structural inversion symmetry and is present
in a variety of bulk chiral magnets.
\cite{muhlbauer09,yu10,tokunaga15,kezsmarki15,seki12} Interfacial DMI
induced by the symmetry breaking in ferromagnetic heterostructures
with asymmetric interfaces \cite{bode07,heide09,yang15} also leads to
the formation of skyrmions in thin ferromagnetic layers
\cite{heinze11,romming13,herve18} and
multilayers.\cite{moreau-luchaire16} Another classical energy term
known to favor non-collinear spin alignment in ferromagnets is the
dipolar energy, also called stray field or demagnetizing
energy.\cite{hubert} A manifestation of the long-range nature of this
energy in thin films with perpendicular magnetic anisotropy is the
appearance of micron-sized magnetic bubble
domains.\cite{kooy60,bobeck67,chaudhari73,montoya17} The equilibrium
shape and size of these domains is determined by the balance between
the long-range dipolar interaction and the energy cost of the bubble
related to the domain wall energy and the Zeeman energy.
\cite{bernand-mantel18}

The orthodox theory of skyrmions in ultrathin ferromagnetic layers
with interfacial DMI relies on a model that accounts for the dipolar
interaction through an effective anisotropy term, neglecting
long-range effects.\cite{ivanov90,rohart13} At the same time, in
single ferromagnetic layers with interfacial DMI, large chiral
skyrmions, also called skyrmionic bubbles have been observed,
\cite{jiang15,buttner15,woo16,yu16a,schott17} suggesting a nontrivial
interplay between DMI and long-range dipolar effects.  The competition
between these two energies also leads to the formation of skyrmions
exhibiting spin rotations with intermediate angles between N\'eel and
Bloch,\cite{buttner18,legrand18,dovzhenko18} a phenomenon also
present in domain walls.\cite{thiaville12} In addition, there is a
growing body of theoretical evidence that points to a need to take
into account the long-range dipolar energy in the models describing
magnetic
skyrmions.\cite{kiselev11,bernand-mantel18,buttner18,legrand18a} In
particular, B\"uttner et al. \cite{buttner18} used a 360$^\circ$-wall
ansatz \cite{braun94a} to numerically calculate the skyrmion
equilibrium radius and energy as functions of the material parameters
for intermediate thicknesses, focusing on room temperature stable
skyrmions and predicting the existence of room temperature skyrmions
stabilized solely by the stray field.

The above considerations put into question the validity of the
commonly used assumption that the contribution of the long-range
dipolar interaction is negligible. Another open question is whether
there exists a size difference between the skyrmions stabilized by the
DMI and those stabilized by the stray
field.\cite{kiselev11,bernand-mantel18,buttner18,legrand18a} In this
paper, we address these questions using an ansatz-free analysis of a
micromagnetic model that is valid for sufficiently small film
thicknesses.  We provide explicit analytical expressions for the
skyrmion radius, rotation angle and energy valid in the low DMI and
thickness regime, taking into account the long-range dipolar energy
contribution. We obtain a prediction for the critical DMI value at
which the skyrmion character changes from pure N\'eel to a mixed
N\'eel-Bloch type. These findings are corroborated by micromagnetic
simulations.  Our rigorous treatment of the stray field contribution
sheds light on the necessity to tune both the magnetic layer thickness
and the DMI constant to optimise the skyrmion size and stability for
applications.

\section{MODEL}

We consider a ferromagnetic thin film with perpendicular magnetic
anisotropy (PMA) and infinite extent in the plane.  The film is
assumed to be sufficiently thin in order for the magnetization vector
$\m$ to be constant in the direction normal to the film plane. Under
these conditions, the micromagnetic energy \cite{hubert} reduces to:
\cite{ms:prsla16,kmn:arma19,m:cvar19}
\begin{align}
  \label{E}
  \begin{split}
    E(\m) & = \int_{\R^2} \left( |\nabla \m|^2 + (Q - 1) |\mpe|^2
    \right) d^2 r \\ & + \kappa \int_{\R^2} \left( \mpa \nabla \cdot
      \mpe -
      \mpe \cdot \nabla \mpa \right) \, d^2 r  \\
    & - {\delta \over 8 \pi} \int_{\R^2} \int_{\R^2} {(\mpa(\mathbf r)
      - \mpa(\mathbf r'))^2 \over |\mathbf r - \mathbf r'|^3} \, d^2 r
    \, d^2 r' \\ & + {\delta \over 4 \pi} \int_{\R^2} \int_{\R^2}
    {\nabla \cdot \mpe(\mathbf r) \, \nabla \cdot \mpe (\mathbf r')
      \over | \mathbf r - \mathbf r'|} \, d^2 r \, d^2 r'.
  \end{split}
\end{align}
Here $E$ is measured in the units of $Ad$, where $A$ is the exchange
stiffness and $d$ is the film thickness, lengths are measured in the
units of the exchange length
$\ell_\mathrm{ex} = \sqrt{2A /(\mu_0 M_\mathrm{s}^2)}$, $M_\mathrm{s}$
is the saturation magnetization and
$\delta = d / \ell_\mathrm{ex} \lesssim 1$ is the dimensionless film
thickness (for further details, see \cite{suppl}). Furthermore, in
Eq.~\eqref{E} we set $\m = (\mpe, \mpa)$, where $\mpe \in \R^2$ and
$\mpa \in \R$ are the respective in-plane and out-of-plane components
of $\m$, and introduced the dimensionless quality factor
$Q = {K_{\mathrm{u}} / K_{\mathrm{d}} }$, where $K_\mathrm{u}$ is the
magnetocrystalline anisotropy constant,
$K_{\mathrm{d}}= \frac12 \mu_0 M_\mathrm{s}^2$, and the dimensionless
DMI strength $\kappa = D / \sqrt{AK_{\mathrm{d}}}$. The first three
energy terms are local and represent, respectively, the exchange
energy, the effective anisotropy energy, which corresponds to the
magnetocrystalline energy renormalized to take into account the local
stray field contribution, and the DMI energy. The last two terms
correspond to the long-range part of the dipolar energy, which splits
into two contributions. The first contribution is due to the
out-of-plane component of $\m$ and accounts for surface charges at the
top and bottom interfaces of the film.  The second energy term
corresponds to volume charges and is due to the in-plane divergence of
the magnetization.

\section{RESULTS AND DISCUSSION}

\subsection{Definition of a skyrmion}
Magnetic skyrmions were originally predicted to exist using a fully
local micromagnetic model that is obtained from Eq.~\eqref{E} by
setting $\delta = 0$.\cite{bogdanov89,bogdanov89a,bogdanov94} Within
this model,\cite {winter61,gioia97} the ground state for PMA
materials ($Q > 1$) and sufficiently small values of $|\kappa|$ is the
monodomain state $\m = \pm \hat{\mathbf z}$, where $\hat{\mathbf z}$
is the unit normal vector to the film plane (the $xy$-plane).
\cite{ms:prsla16,bogdanov94} Therefore, one should identify magnetic
skyrmions with metastable magnetization configurations that locally
minimize the energy in Eq.~\eqref{E}. In addition, skyrmions possess a
non-zero topological charge $q \in \mathbb Z$ defined as
\cite{nagaosa13,braun12,rybakov19}
\begin{align}
  \label{q}
  q(\m) = {1 \over 4 \pi} \int_{\R^2} \m \cdot \left(
  {\partial \m \over \partial x} \times {\partial \m
  \over \partial y} \right) \, dx \, dy, 
\end{align}
provided
$ \lim_{|\mathbf r| \to \infty} \m(\mathbf r) = -\hat{\mathbf z}$ in
order to fix the sign convention so that $q = +1$ for either the
N\'eel or Bloch skyrmion profiles.  In a fully local micromagnetic
model with bulk DMI and no anisotropy term, Melcher \cite{melcher14}
studied the existence of minimizers among nontrivial topological
sectors in the presence of a sufficiently strong Zeeman term. He found
that in this class the energy is globally minimized by a configuration
with $q = +1$ (in our convention) and identified this energy
minimizing magnetization configuration with a magnetic skyrmion.

In contrast, in the absence of an applied magnetic field and in the
presence of long-range dipolar interaction the monodomain state is
never the ground state in an extended ferromagnetic
film.\cite{kaplan93,kmn:arma19} This can be seen by noting that the
energy in Eq.~\eqref{E} with $Q > 1$ and $\delta > 0$ goes to negative
infinity for configurations consisting of a growing bubble domain in
which the $\m = +\hat{\mathbf z}$ core is separated from the
$\m = -\hat{\mathbf z}$ background by a Bloch or N\'eel wall,
depending on the magnitude of $|\kappa|$, and which carry the
topological charge $q = +1$.\cite{kmn:arma19,bernand-mantel18} Thus,
it is not possible to carry out the analysis of
Ref.~[\onlinecite{melcher14}] to establish existence of skyrmion
profiles via direct energy minimization without introducing further
restrictions on the admissible configurations distinguishing compact
magnetic skyrmions from skyrmionic bubbles.

In the present work, we assign a mathematical meaning to the notion of
compact magnetic skyrmion by defining a class of admissible
configurations in which the topological charge is fixed to $q = +1$
and the exchange energy cannot exceed twice the topological lower
bound, i.e., twice the exchange energy of the Belavin-Polyakov
profile.\cite{belavin75,buttner18,bernand-mantel18} Within this class
we establish the existence of compact skyrmions as minimizers of the
energy in Eq.~\eqref{E} for  an explicit range of the parameters.
  \begin{theorem}
      Let $Q > 1$, $\delta > 0$ and $\kappa \in \R$ be such that
      \begin{align}
        \label{condQd}
        (2 |\kappa| + \delta)^2 < 2 (Q - 1).
      \end{align}
      Then there exists a minimizer of $E$ among all $\mathbf m$ such
      that $q(\mathbf m) = +1$,
      $\int_{\R^2} |\nabla \mathbf m|^2 d^2 r < 16 \pi$, and
      $\mathbf m(\mathbf r) \to -\hat{\mathbf{z}}$ as
      $|\mathbf r| \to \infty$.
  \end{theorem}
  A more precise statement and a sketch of the proof of this result
  may be found in the Supplemental Material.\cite{suppl}  In the
  following, we always refer to the minimizers in the above theorem as
  {\em skyrmion solutions.} We note that Eq.~\eqref{condQd} is only a
  sufficient condition for their existence.

\begin{figure*}
  \centering
  \includegraphics[width=15cm]{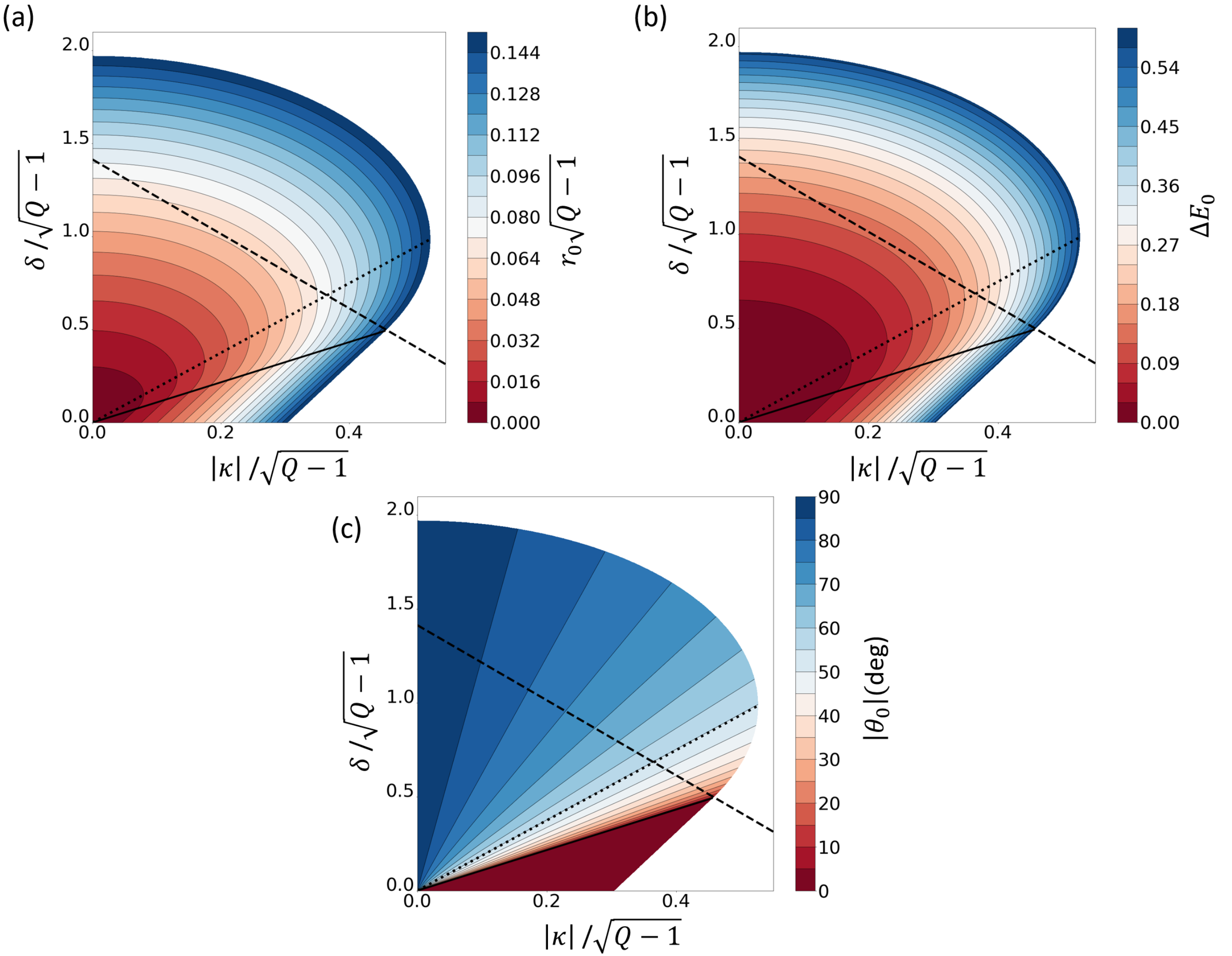}
  \caption{Dependences of the skyrmion characteristics on the
    parameters obtained from the asymptotic analysis for
      $|\kappa|, \delta \ll \sqrt{Q - 1}$: (a) the dimensionless
    skyrmion radius $r_0$ from Eq.~\eqref{lambda0}; (b) the
    skyrmion collapse energy barrier $\Delta E_0$ from
      Eq.~\eqref{E0}; and (c) the rotation angle $|\theta_0|$
    from Eq.~\eqref{theta0}. 
    The solid line shows the transition from N\'eel to mixed
      N\'eel-Bloch skyrmions governed by Eq.~\eqref{treshDsky}. The
      dashed line corresponds to the boundary of the region defined in
      Eq.~\eqref{condQd} in which existence of skyrmion solutions is
      guaranteed. The dotted line shows the parameters at which the
      skyrmion radius achieves its minimum value as a function of
      $\delta$ according to Eq.~\eqref{deltamin}.  }
  \label{solutions}
\end{figure*}

\subsection{\label{subsec:sol}Skyrmion solutions in the low $|\kappa|$
  and $\delta$ regime}

Once existence of a compact skyrmion solution is established, we
proceed with an asymptotic analysis of the skyrmion profile at low
values of $|\kappa|$ and $\delta$ (for a fully rigorous treatment, see
\cite{bms}). It is known that in a model with exchange energy alone
the minimizer is given explicitly by all rigid in-plane rotations,
dilations and translations of the Belavin-Polyakov
profile.\cite{belavin75} It can be expected that in the limit where
additional energy terms appear as perturbations (low $|\kappa|$ and
$\delta$) of the dominating exchange energy, skyrmions retain the
Belavin-Polyakov profile.\cite{abanov98} This has been demonstrated
recently via a formal asymptotic analysis of radial skyrmion solutions
in the local model with bulk DMI in the limit of vanishing DMI
constant.\cite{komineas19} In the full model given in Eq.~\eqref{E},
we were able to prove \cite{bms} that as $\kappa,\delta \to 0$ the
energy minimizing profile $\m$ converges to a Belavin-Polyakov profile
$\m_0$ given by
\begin{align}
  \label{BP}
  \m_0(\mathbf r) = \left( -{2 r_0  R_{\theta_0} \mathbf r
  \over  |\mathbf  r|^2 + r_0^2
  }, {   r_0^2-|\mathbf r|^2 
  \over  r_0^2 + |\mathbf r|^2 } \right) \qquad \mathbf r \in \mathbb
  R^2, 
\end{align}
where $R_{\theta_0}$ is the $2 \times 2$ matrix of in-plane rotations by
angle 
\begin{align}
  \label{theta0}
  \theta_0 =
  \begin{cases}
    0 & \text{ if } \kappa \geq {3 \pi^2
      \over 32} \delta,\\
    -\pi & \text{ if } \kappa \leq -{3 \pi^2
      \over 32} \delta,\\
    \pm \arccos\left(\frac{32 \kappa}{3 \pi^2 \delta} \right) &
    \text{ else,}
  \end{cases}
\end{align}
and the dimensionless skyrmion radius is
asymptotically
\begin{align}
  \label{lambda0}
  r_0 \simeq \frac{1}{16\pi \sqrt{Q - 1}} \times
  \frac{\bar\eps(\kappa,\delta,Q)}{|\log
  \left(\beta\bar\eps(\kappa,\delta,Q)\right)  |},  
\end{align}
for $\beta \bar \eps \ll 1$ with $\beta \approx 0.04816$ and
\begin{align}\label{epsilon_definition}
  \bar \eps(\kappa,\delta,Q) = {1 \over \sqrt{Q - 1}} \times
  \begin{cases}
    \left(8\pi |\kappa| - {\pi^3 \over 4} \delta \right) & \text{ if
    }|\kappa| \geq {3 \pi^2
      \over 32} \delta,\\ 
    \left(\frac{128 \kappa^2}{3 \pi
        \delta} + {\pi^3 \over 8} \delta \right) & \text{ else.}
  \end{cases}
\end{align}
The above expressions may be obtained by considering a suitably
truncated magnetization profile in the form of Eq.~\eqref{BP},
optimizing in $\theta_0$ and $r_0$ and expanding the obtained
expressions in the leading order of $\delta$ and $|\kappa|$ (see
\cite{suppl} for more details). Our analysis also yields the following
asymptotic expression for the skyrmion energy:
\begin{align}
  \label{E0}
  E_0 \simeq 8 \pi - 
  \frac{\bar \eps ^2 (\kappa,\delta,Q)}{32\pi
  |\log\left(\beta\bar \eps(\kappa,\delta,Q)\right)|}. 
\end{align}
The associated skyrmion collapse energy barrier
$\Delta E_0 = 8 \pi - E_0$ gives an indication of the skyrmion
stability as it represents the energy necessary to suppress the
skyrmion via
compression.\cite{cortes-ortuno17,buttner18,bernand-mantel18} The
solution described in Eqs.~\eqref{BP}--\eqref{E0} is asymptotically
exact to the leading order for $| \kappa| \ll \sqrt{Q - 1}$ and
$ \delta \ll \sqrt{Q - 1}$, but in practice remains at least
qualitatively correct also up to $ \kappa \sim \sqrt{Q -1}$ and
$\delta \sim \sqrt{Q - 1}$. Nevertheless, to avoid artefacts from
  the predictions of our formulas outside their range of validity, we
  somewhat arbitrarily restrict the considered parameters to those for
  which $\beta \bar\eps(\kappa, \delta, Q) \leq e^{-1}$, ensuring
  $|\log (\beta \bar\eps(\kappa, \delta, Q))| \geq 1$.

\subsection{\label{subsec:dep}Asymptotic properties of skyrmion
    solutions}

The dependences of the dimensionless skyrmion radius $r_0$, the
collapse energy $\Delta E_0$ and the rotation angle $\theta_0$ on the
model parameters obtained from the asymptotic analysis of
  Sec.~\ref{subsec:sol} are presented in Fig.~\ref{solutions}.  The
first important characteristic of the solution is the existence of a
minimum or threshold $|\kappa|$ value
\begin{align}
  \label{treshDsky}
  |\kappa|_{\mathrm{sky}}^{\mathrm{thresh}}=\frac{3\pi^2}{32} \,
  \delta, 
\end{align}
above which pure N\'eel skyrmions ($\theta_0=0$ or
  $\theta_0 =-\pi$, depending on the sign of $\kappa$) are obtained
and below which skyrmions are characterized by a non-zero rotation
angle $\theta_0$. This angle tends to $\pm \pi/2$, corresponding to
pure Bloch skyrmions when $\kappa \to 0$. It is a direct consequence
of the competition between long-range dipolar interaction, which
favors a Bloch rotation, and interfacial DMI, which favors a N\'eel
rotation. Note that a similar threshold is observed in the case of
straight domain walls:\cite{thiaville12,lemesh17}
\begin{align}
  \label{treshDwall}
  |\kappa|_{\mathrm{wall}}^{\mathrm{thresh}}=\frac{4\ln 2}{\pi^2} \, \delta.
\end{align}
 Thus, a larger DMI is necessary to obtain a pure N\'eel
skyrmion as compared to the case of a 1D N\'eel wall, as can be seen
from the factor of $\sim$ 3 difference between the values of
$|\kappa|_{\mathrm{sky}}^{\mathrm{thresh}}$ and
$ |\kappa|_{\mathrm{wall}}^{\mathrm{thresh}}$. This is an indication
that dipolar effects play a stronger role for skyrmions compared to
domain walls.

The second characteristic associated with the interplay between DMI
and the dipolar interaction that is visible in
Fig.~\ref{solutions}(a) and Fig.~\ref{solutions}(b) is the
non-monotone dependence of the dimensionless skyrmion radius $r_0$ and
collapse energy $\Delta E_0$ on $\delta$ for $Q$ and $\kappa$
fixed. For $\delta$ below the critical value where skyrmions are of
N\'eel character, the skyrmion radius decreases with increasing
$\delta$, while for large enough $\delta$, in the regime with non-zero
$\theta_0$, the radius increases with $\delta$.  As can be seen
  from Eq.~\eqref{lambda0}, the skyrmion radius reaches its minimum at
  $\delta = \delta_\mathrm{opt}$, where
  \begin{align}
    \label{deltamin}
    \delta_\mathrm{opt} = {32 |\kappa| \over \pi^2 \sqrt{3}}. 
  \end{align}
This observation is of importance for applications, as the thickness
of the film is the parameter which is the most easy to tune
experimentally for a thin film in order to optimize the skyrmion size
and stability. Interestingly, at $\delta = \delta_\mathrm{opt}$
  the rotation angle $\theta_0$ attains a universal value of
  $\theta_0^\mathrm{opt} = \pm
  \arccos(\mathrm{sgn}(\kappa)/\sqrt{3})$, i.e.,
  $\theta_0^\mathrm{opt} \approx \pm 54.74^\circ$ for $\kappa > 0$ or
  $\theta_0^\mathrm{opt} \approx \pm 125.3^\circ$ for $\kappa < 0$.

The third important result illustrated in Fig.~\ref{solutions} is the
existence of skyrmions stabilized solely by the long-range dipolar
interaction for $\kappa = 0$. Such dipolar skyrmions possess a pure
Bloch character ($\theta_0= \pm \pi/2$), with volume charges not
contributing to the energy.  We observe in Fig.~\ref{solutions}(a)
that, starting from $\kappa = 0$ and following a skyrmion solution of
fixed radius while decreasing $\delta$, one goes continuously from a
Bloch skyrmion at $\kappa = 0$ to a N\'eel skyrmion at $\delta =
0$. Consequently, skyrmions stabilized by DMI and stray field cannot
be distinguished by their radius.

\begin{figure}
  \centering
  \includegraphics[width=7cm]{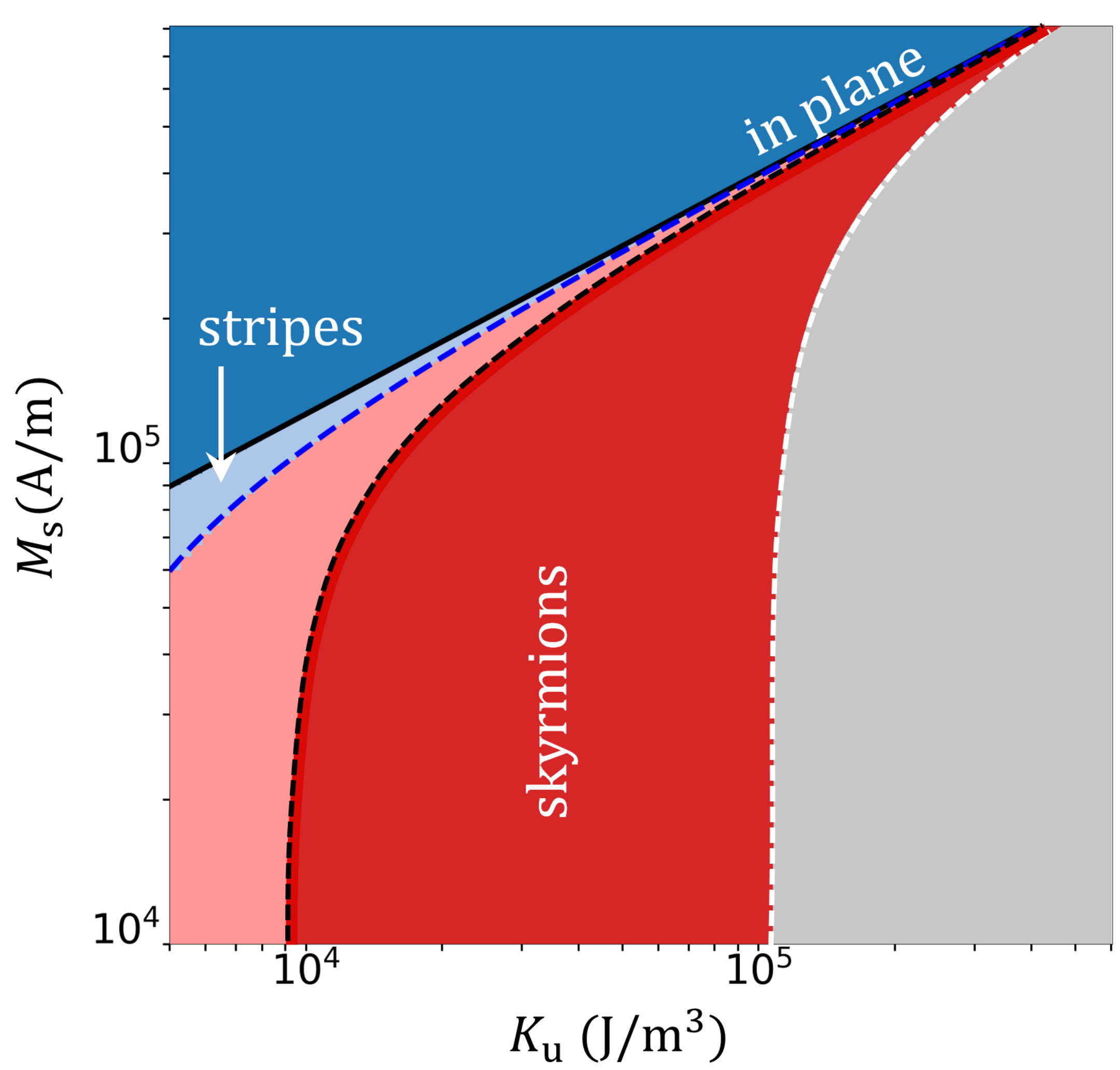}
  \caption{Skyrmion phase diagram for $A=20$ pJ/m, $D = 0.3$
      mJ/m$^2$, and $d = 1$ nm. The dark red zone is the domain of
      existence of our skyrmion solutions (see Sec.~\ref{subsec:phase}
      for a complete description of the different zones and lines). }
  
  \label{diag}
\end{figure}

\begin{figure*}
  \centering \includegraphics[width=14cm]{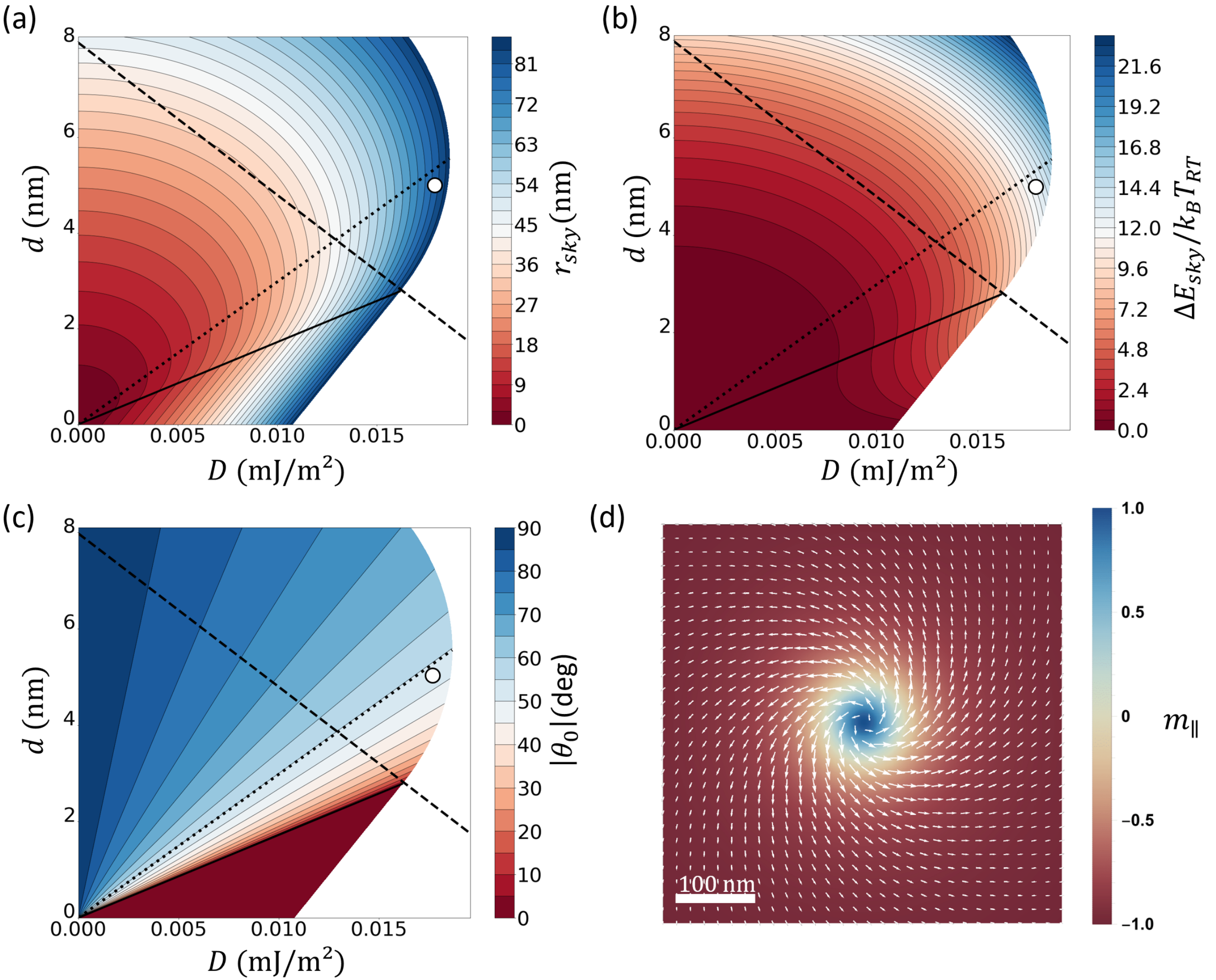}
  \caption{ Dependences of the skyrmion characteristics in the low DMI
    regime.  The parameters are $A=20$ pJ/m,
      $M_{\mathrm{s}}= 10^5$ A/m and $K_\mathrm{u} = 6346$ J/m$^3$
      corresponding to $Q = 1.01$. (a) The skyrmion radius
    $r_{\mathrm{sky}}$. (b) The normalized skyrmion collapse energy
    barrier $\Delta E_\mathrm{sky}
    / (k_{\mathrm{B}}T_{\mathrm{293K}})$. 
    (c) The rotation angle $|\theta_0|$. (d) The skyrmion profile
      obtained numerically for $d=5$ nm and $D=0.018$ mJ/m$^2$,
      corresponding to the white dot in panels (a)--(c), using
      MuMax3\cite{vansteenkiste14} on a $4096 \times 4096$ nm$^2$
      square domain subject to periodic boundary conditions, with the
      mesh size of $4 \times 4 \times 5$ nm$^3$. The image in
      (d) is constructed by superimposing the in-plane magnetization
    $\mpe$ represented with arrows and the out-of-plane magnetization
    $\mpa$ represented by a colormap. The lines in (a)--(c) are
      the same as in Fig. \ref{solutions}.}  
  \label{lowD}
\end{figure*}

\subsection{\label{subsec:phase}Phase diagram}

To complete our description, we locate the skyrmion solutions on a
phase diagram (see Fig.~\ref{diag}). For that purpose, we fix
a representative set of parameters: exchange constant $A = 20$
  pJ/m, film thickness $d=1$ nm and DMI constant $D= 0.3$ mJ/m$^2$,
and vary the saturation magnetization $M_\mathrm{s}$ and
magnetocrystalline anisotropy constant $K_\mathrm{u}$ over a wide
range. The solid black line represents the threshold at which the
magnetization reorientation transition between in-plane and
out-of-plane occurs ($Q = 1$, i.e., for
$K_{\mathrm{u}} =K_\mathrm{d}$).  In the dark blue region above this
line, the magnetization prefers to lie in the film plane, and no
compact skyrmion solutions exist in an infinite film.  Below this
line, the easy axis is perpendicular to the film plane. In the zone
represented in light blue, the domain wall energy density defined as
$\sigma_{\mathrm{wall}}=4\sqrt{A(K_{\mathrm{u}} - K_\mathrm{d})}-\pi
D$ is negative. Here, the expected ground state of the thin film is
the helicoidal state,\cite{bogdanov99} and isolated compact skyrmions do not
exist in the absence of an applied magnetic field \cite{romming13}.
Below the dashed blue line corresponding to
$K_{\mathrm{u}}^{\mathrm{crit}}=\frac{\pi^2D^2}{16A} +
K_{\mathrm{d}}$, the ferromagnetic ground state is restored, the
domain wall energy becomes positive again, and compact skyrmions may
exist as metastable states. In the light red region, the existence
of compact skyrmion solutions as metastable state is not
  guaranteed. Indeed, in this region close to the transition to the
  helicoidal state, skyrmions may be subject to elliptical
  instabilities favored by both the DMI and the long-range dipolar
  interaction.\cite{bogdanov94a,hubert} The dark red zone represents
the domain of existence of our skyrmion solutions. It is delimited on
one side by a dashed black line, which represents the boundary of the
region defined by Eq.~\eqref{condQd} below which we have existence of
a compact skyrmion. When the anisotropy is further increased (or
$M_{\mathrm{s}}$ is decreased), the limit of validity of our 2D thin
film model is reached as the skyrmion radius becomes of the order of
the film thickness.  The white dashed line represents the line at
which $r_{\mathrm{sky}}=d$ as a guide to the eye. We point out that
below this line skyrmion solutions may exist and develop
$z$-dependence.\cite{legrand18,legrand18a} Further below this line the
continuum micromagnetic model is no longer valid, as the skyrmion
radius becomes of the order of the interatomic spacing.

\subsection{Application to specific examples for low and intermediate
  $D$ values} \label{sec:ferri}

In this section, we apply our compact skyrmion results to the case of
{\em ferrimagnetic} materials, i.e., materials with low
$M_{\mathrm{s}}$ and $K_{\mathrm{u}}$ values (e.g.,
GdCo\cite{caretta18}). These conditions favor the observation of
skyrmions in the absence of an applied magnetic field as discussed in
the previous section.  An observation of room temperature zero-field
skyrmions in this material was recently reported.\cite{caretta18}

 \begin{figure*}[t]
  \centering \includegraphics[width=15cm]{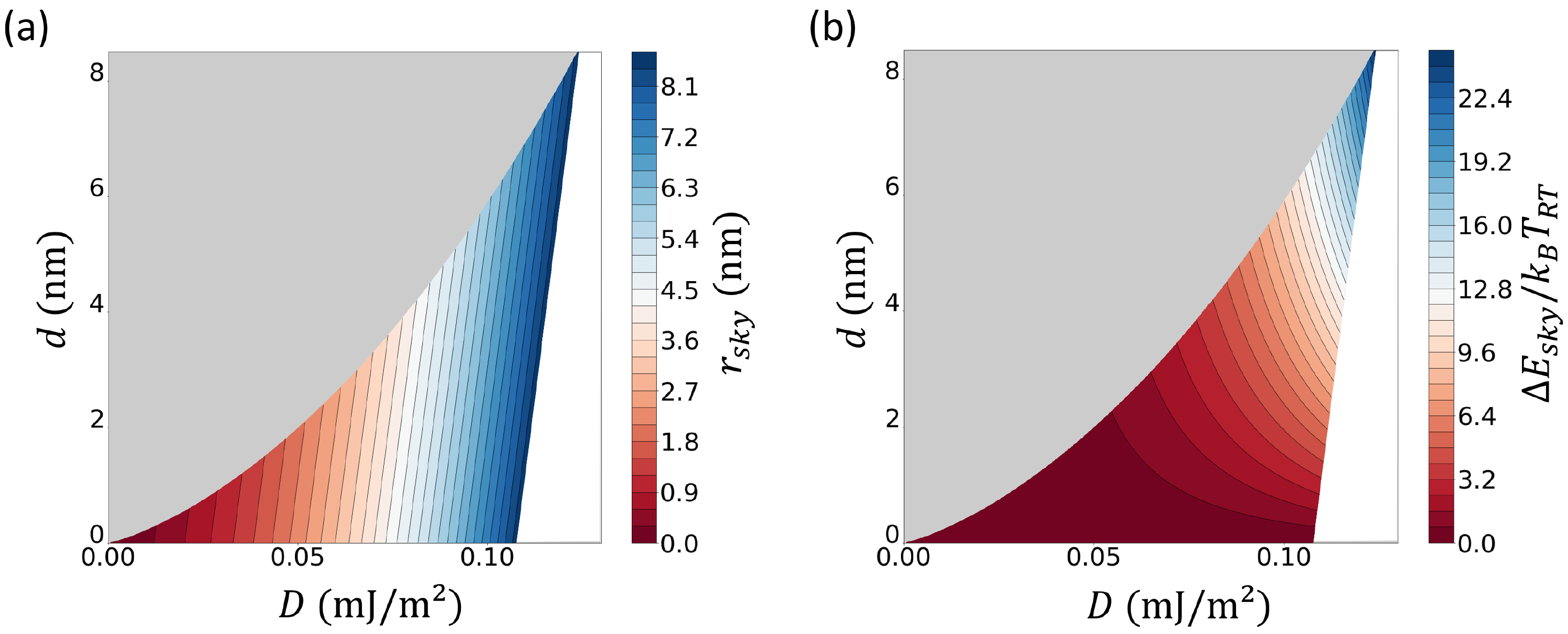}
  \caption{Dependences of the skyrmion characteristics in the
    intermediate DMI regime.  The parameters are $A=20$ pJ/m,
$M_{\mathrm{s}}= 10^5$ A/m and   $K_\mathrm{u} = 1.26 \times 10^4$ J/m$^3$ 
corresponding to $Q = 2$.  (a) The skyrmion radius
    $r_{\mathrm{sky}}$. (b) The normalized skyrmion collapse energy
    barrier $\Delta E_\mathrm{sky} /(k_{\mathrm{B}}T_{\mathrm{293K}})$. 
    The grey zone corresponds to the region where $r_\mathrm{sky} <
    d$. }
  \label{intermD}
\end{figure*}

In Fig.~\ref{lowD}, we present the case of low $D$ values for which
the transition from pure N\'eel to pure Bloch skyrmion appears, as
seen in Fig.~\ref{lowD}(c).  For comparison, we carried out
micromagnetic simulations for the parameters corresponding to the
white dot in Fig.~\ref{lowD}(a)--(c) (see figure caption for
  details). From the simulations we obtain a skyrmion with a rotation
angle $\theta_0\simeq 46^\circ$ and a radius
$r_\mathrm{sky} \simeq 52$ nm, vs.\ $\theta_0\simeq 52^\circ$ and
$r_\mathrm{sky} \simeq 80$ nm from the analytical formulas.  This
confirms the transition from purely N\'eel, to intermediate
N\'eel-Bloch rotation angle, predicted by our analysis at lower
thicknesses compared to 1D walls as discussed in Sec.~\ref{subsec:dep}
.  Indeed, for 1D walls, a purely N\'eel character is expected up to
thicknesses of $\sim 10$ nm.  This validates the increased importance
of dipolar interaction in the case of compact skyrmions predicted by
our theory compared to 1D walls.  Figure~\ref{lowD}(b) shows the
skyrmion collapse energy barrier
$\Delta E_\mathrm{sky} = \Delta E_0 Ad$ normalized by the room
temperature thermal energy $k_{\mathrm{B}}T_{\mathrm{293K}}$. The
collapse energy barrier $\Delta E_\mathrm{sky}$ obtained by our
analysis is 13 $k_{\mathrm{B}}T_{\mathrm{293K}}$, compared to
15.5 $k_{\mathrm{B}}T_{\mathrm{293K}}$ from the micromagnetic
simulations.  We observe that in the low $D$ regime the collapse
barrier increase with film thickness.  This consideration justifies
the choice of systems with bulk out-of-plane anisotropy (like the
ferrimagnetic alloy GdCo), or multilayers (e.g., (Pt/Co/Ir)$_n$) to
optimise skyrmion lifetime, since they allow to increase the film
thickness (or effective thickness) without losing the out-of-plane
anisotropy, as would be the case for single ferromagnetic layers with
surface-induced anisotropy alone.\cite{caretta18,moreau-luchaire16}

In Fig.~\ref{intermD}, we present the results for an intermediate DMI
range where the $D$ values are an order of magnitude larger than those
in Fig.~\ref{lowD}. We use the same parameters as in Fig.~\ref{lowD},
except $K_\mathrm{u} = 1.26 \times 10^4$ J/m$^3$ corresponding to
$Q = 2$.  All the solutions in Fig.~\ref{intermD} are the iconic
N\'eel skyrmions with $\lesssim 10$ nm radii that grow with an
increase of the DMI strength [see Fig.~\ref{intermD}(a)]. The skyrmion
collapse barrier can be heightened by either increasing the film
thickness or the DMI strength [see Fig.~\ref{intermD}(b)].  In the
  low thickness regime, the decrease of the collapse barrier with the
  film thickness is due to the dimensional scale factor of $Ad$. The
  same phenomenon is at the origin of the short skyrmion lifetime
  ($\sim 1s$) observed experimentally at low temperature in
  ferromagnetic monolayers,\cite{romming13} despite a large DMI
  constant.\cite{romming15}
  
We observe in Fig.~\ref{intermD}(b) that the collapse energy barrier
of $\sim $ 8 nm radius skyrmions reaches 22 $k_{\mathrm{B}}T_\mathrm{293K}$,
which corresponds to a few seconds lifetime, considering the
N\'eel-Brown model with the attempt frequency $\nu = 10^{9}$ s$^{-1}$.
At fixed thickness $d$, the skyrmion radius decreases with the DMI
strength, and the limit of validity of our thin film model is reached
as the skyrmion radius becomes of the same order as the film
thickness. In this regime, 3D models and full 3D micromagnetic
simulations will be needed to take into account the long-range dipolar
effects.

\section{Summary}

We have used rigorous mathematical analysis to develop a skyrmion
theory that takes into account the full dipolar energy in the thin
film regime and provides analytical formulas for compact skyrmion
radius, rotation angle and energy. While long-range interactions are
often assumed to have a negligible impact on skyrmions in this regime,
we demonstrate that the DMI threshold at which a compact skyrmion
looses its N\'eel character is a factor of $\sim3$ higher than that
for a 1D wall. A reorientation of the skyrmion rotation angle from
N\'eel to intermediate N\'eel-Bloch angles is predicted as the layer
thickness is increased in the low DMI regime, which is confirmed by
micromagnetic simulations. The estimation of this reorientation
thickness is important for applications as the skyrmion angle affects
its current-induced dynamics.\cite{tomasello14}\\
 
 \begin{acknowledgements}
   
   A.\ B.-M. wishes to acknowledge support from DARPA TEE program
   through grant MIPR\# HR0011831554. The work of C.\ B.\ M. and T.\
   M.\ S. was supported, in part, by NSF via grants DMS-1614948 and
   DMS-1908709. C.\ B.\ M.  would also like to acknowledge support
     by CNRS via EUR grant NanoX ANR-17-EURE-0009 in the framework of
     the ``Programme des Investissements d'Avenir'', and LPCNO, INSA,
     Toulouse, France, for its hospitality. This work benefited from
     access to the HPC resources of CALMIP supercomputing center under
     the allocation 2019-19011. The authors also acknowledge the
     hospitality of Erwin Schr\"odinfer International Institute for
     Mathematics and Physics.

 \end{acknowledgements}


\end{document}